# Brush structures directly anchored to ion beam treated polymer surfaces without linker

Alexey Kondyurin

*School of Physics, University of Sydney, Australia*
*Leibniz Institute of Polymer Research Dresden, Dresden, Germany*
*Ewingar Scientific, Ewingar, Australia*

**Abstract**

A brush structure is an interesting object for future applications in medical and electronic devices. Usual substrate for the brushes is silicon wafer with linker molecules. In present study an ion beam treatment of polymer was used for attachment of brush structures without liker molecules. The goal of the study was a fabrication of carbonized active substrate and a direct attachment of different brush molecules. The carbonised coating on silicon wafer has been prepared from ion beam implanted polystyrene coating and characterised with AFM, XPS, ESR, Raman, FTIR and ellipsometry measurements. The brush structures based on polystyrene and polyacrylamide backbone with thiol, amine and carboxyl end groups have been synthesised on the carbonised substrates. The brush structures have been characterised with ellipsometry, XPS, FTIR and AFM. The swollen brush shows a thickness phase transition with temperature rise. The attachment of the brush structures is based on free radical reactions on the carbonised surface. The effect of impurity of the brush polymer was found significant. Thus, the brush structures of broad kinds of the end groups can be synthesised on the carbonised substrates without a linker.

**1. Introduction**

In recent years, there has been an interest surrounding the modification of surfaces for producing their affinity with external moieties such as inorganic nanoparticles [1] and brush molecules [2], where the surface area is critical to their performance such as antifouling [3], thermosensor [4], nanosensors [5], bio sensors [6], chemical sensor [7], adsorbents [8], friction agent [9] etc. Polymer brushes have been used to render a variety of functionalities including manageable adsorption of protein molecules [10-14]. Attached protein and DNA have become useful in a range of applications such as next-generation sequencing, hybridization arrays, protein biosynthesis compartments, and coated particle assembly [15]. Polymer brushes are the systems in which polymer chains are grafted on the surfaces or interfaces by one chain end with an average distance between two anchoring points significantly smaller than the radius of gyration of similar polymer chains floating free in solution. Usually, the attachment of brush molecules and nanoparticles needs a linker, which must provide a specific chemical activity to the substrate and to brush molecule simultaneously [16]. It requires a special chemical activity of the substrate material and brush molecules. It limits a range of application of the brush structures. A universal substrate activity could extend the field of brush structure application.

In the present study the attachment of the brush molecules to the highly active carbon surface layer of polymer is proposed (Fig.1). The carbon layer is generated by high-energy ions which penetrate into the carbon-rich polymer and break the chemical bond in the macromolecules [17]. The restructuring of the chemical bonds after breakage gives the carbonized structures like amorphous carbon and graphitic clusters. The universal chemical activity of carbon layer is provided by free radicals on the edge of graphitic clusters. These transformations are observed for most kinds of polymers with sufficient concentration of the carbon atoms in a pristine macromolecule: polyethylene, polypropylene, polystyrene and others [17]. An ability of the carbonized layer after ion beam implantation to attach the brush molecules and the properties of the brush structures on the carbonised substrate is an aim of the present investigation.

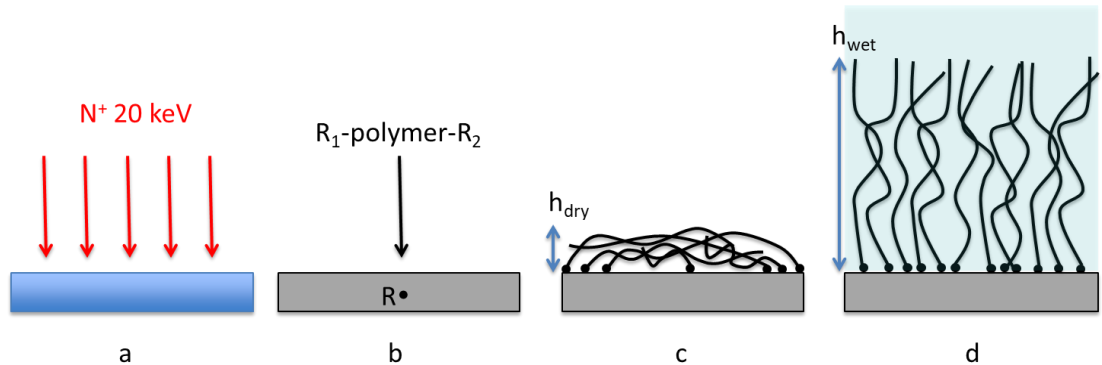

Fig.1. Molecular brush formation on carbonized surface. (a) First stage is ion beam treatment of the polymer substrate to create the active carbon surface. (b) Second stage is covalent attachment of brush molecules with $R_1$ and $R_2$ functional groups. Not attached brush molecules are washed out. (c) Collapsed dry brush layer. $h_{dry}$ is a thickness of the dry brush layer. (d) Standing brush layer in $\Theta$-solvent. $h_{wet}$ is a thickness of the swollen brush layer.

## 2. Experiment

Polystyrene (PS) films of 70 nm nominal thickness were prepared by spin coating onto (100) silicon substrates (square of 10x10 mm) at 2000 rpm using a SCS G3P-8 Spincoater. The spin coating solution consisted of polystyrene (Austrex 400 from Polystyrene Australia Pty. Ltd) dissolved to a concentration of 5 g/l in toluene (Sigma Aldrich, Australia, Product no. 34866, purity >99.9%). The solution's concentration and spin coating rate were selected to give a homogenous film thickness over the entire silicon wafer. The presence of toluene in the spincoated PS films was monitored by FTIR spectroscopy.

An inductively coupled radio-frequency (13.56 MHz) plasma was used as the source for plasma immersion ion implantation (PIII). The base pressure of the vacuum chamber was $5 \cdot 10^{-5}$ Torr. The pressure of nitrogen during implantation was $2.0 \cdot 10^{-3}$ Torr. The plasma power was 100 W with reverse power of 15 W when matched. Acceleration of ions from the plasma was achieved by the application of high voltage 20 kV bias pulses of 20 μs duration to the sample holder at a frequency of 50 Hz.

The silicon wafers coated with polystyrene film were mounted on a metal substrate holder. Ion implantation occurred through a metal grid which was electrically connected to the holder and held parallel to the sample, 5 cm in front of it. The samples were treated for durations of 800 sec, corresponding to implantation ion fluences of $1 \times 10^{16}$ ions/cm². For some measurements, shorter treatment time and lower ion energies were used to get a trend of the PS properties. In such cases, the variable parameters are noted at the figure legends. After the treatment the samples were stored in sealed boxes protected against dust in darkness at room temperature. The treated surface was kept away from any contact with tweezer, wall of storage box or other materials as well as breath of staff. The treated samples were stored 5 weeks before the brush polymers were attached. In some experiments the samples were stored 3 weeks as it is mentioned additionally in the results section.

Ion fluence estimates were obtained from the number of high voltage pulses multiplied by the fluence corresponding to one pulse. The fluence corresponding to one high voltage pulse (with bias 20 kV) was estimated in previous experiments on polyethylene films treated using the same PIII process. The fluence in these experiments was calibrated by comparing the UV transmission spectra of the treated polyethylene films and with data obtained in previous PIII and ion beam treatments with 20 keV ions and known fluences.

The thicknesses and optical constants of the spun polystyrene films, before and after exposure to the PIII treatment, were determined using a Woollam M2000V (in University of Sydney, Sydney, Australia) and Woollam M2000VI (in Institute of Polymer Research, Dresden, Germany) spectroscopic ellipsometers. Ellipsometric data was collected for five angles of incidence: 55, 60, 65, 70 and 75°. In the case of the untreated film, a model consisting of a transparent Cauchy layer on top of the silicon substrate was sufficient to achieve a good fit to the data. When fitting the data collected for PIII treated samples, a model with a transparent Cauchy layer in 600-1500 nm region of wavelength was attempted initially. If a good fit was not obtained for all five angles of incidence, then an absorption factor was added to the Cauchy layer in the model.

Transmission FTIR spectra of the carbonized PS on the silicon wafer were recorded before and after plasma treatment using a Bomem FTIR spectrometer. The spectral resolution was 4 $cm^{-1}$, number of scans is 500. The spectrum of silicon wafer was subtracted. The optical density of spectral lines associated with particular bond vibrations were used to analyse the residual PS structures in the carbonised coating.

Micro-Raman spectra ($\lambda$=532.14 nm) were obtained in the backscattering mode using a diffraction double monochromator spectrometer HR800, Jobin Yvon, LabRam System 010. The spectral resolution was 4 $cm^{-1}$ and the number of scans and integration time were varied to ensure sufficient signal-to-noise ratio. An optical microscope (Olympus BX40) with a ×100 objective was used to focus the laser beam and collect the scattered light. LabRam software was used to analyse the spectra.

The wettability of the carbonized substrate was measured using the sessile drop method. Kruss equipment DS10 was employed to measure the contact angles of mQ-water.

The electron spin resonance spectra of free radicals in the PIII treated PS was investigated by means of an electron spin resonance (ESR) spectrometer, Bruker Elexsys E500, operating in X band with a microwave frequency of 9.35 GHz and a central magnetic field of 3330 G at $20^0$C. The spectrometer was calibrated using a pitch sample with α,α'-diphenyl-β-picrylhydrazyl. As a control, the ESR spectra of empty tubes and untreated PS were measured.

The brush polymers were purchased from Polymer Source, Inc. Canada (Table). Two kinds of the brush polymers were selected: based on polystyrene chain and on N-isopropylacrylamide (NIPAM) chain. A molecular mass of the polymers varied from 3000 to 553000 (Table). The end groups were selected for different activity to free radicals: amine, thiol and carboxylic. Each polymer was attached and measured on 3 carbonized samples for statistics.

The tetrahydrofurane (THF) was purchased from Sigma-Aldrich. A thin film of the brush polymer dissolved in THF was spin-coated and annealed at 120 °C (for PS based brushes) or at 150 °C (for NIPAM based brushes) for 18 h in a vacuum oven. To ensure the removal of unattached polymer molecules from the substrates, the samples were cleaned in THF for 2 h in Soxhlet and dried.

In the case of attachment from solution, the drop of the solution was placed on the substrate and cover with two caps against evaporation of the solvent. To hold the constant vapour pressure of the solvent over the solution drop, some drops of pure solvent were placed under $1^{st}$ and $2^{nd}$ caps. The substrate with wet drop was kept 17h at room temperature. The drop was not dried after 17h of the storage. Then the substrate was washed as described above.

Transmission FTIR spectra of the brush samples were recorded using a Bruker FTIR spectrometer. The spectral resolution was 4 $cm^{-1}$, number of scans was 4000 that provided sufficient signal/noise ratio. The spectra of silicon wafer and carbonized coating have been subtracted. In some case, when it was needed, the water vapour spectra have been subtracted. The final spectra were corrected with baseline and smoothed.

The thicknesses and optical constants of the brush layer were determined using a Woollam M2000VI spectroscopic ellipsometer. The ellipsometric data was collected for five angles of incidence: 55, 60, 65, 70 and 75°. A model consisting of a transparent Cauchy layer on top of the silicon substrate with the carbonised layer measured before was sufficient to achieve a good fit to the data in the 600-1500 nm region of wavelength. The ellipsometry data were used to estimate a brush structure of the grafted polymer. The grafting density of the brush layer was calculated as

$$\sigma = \frac{d\rho N_A}{M_n} \quad (1)$$

where d is thickness of a dry brush layer, ρ is a brush density, $N_A$ is Avogadro's constant, $M_n$ is a number average molecular weight of a brush [2].

The reduced tethered density was calculated as

$$\Sigma = \sigma \pi R_g^2 \quad (2)$$

where $R_g$ is the radius of gyration of the brush molecule. A classification of the brush kind was based on the tethered density calculation with following assumptions: a grafted polymer with Σ>5 is a brush structure, a grafted polymer with 1<Σ<5 is a transition regime, and a grafted polymer with Σ<1 is a mushroom structure.

The thickness and optical constants of the swollen brush layer were determined using a Woollam M2000VI spectroscopic ellipsometer equipped liquid cell Gerber Instruments (Kuevette 770.044-QG, Quarzglas Spectrosil 2000). The ellipsometric data was collected for 70° angle of incidence.

The surface analysis of brush structures was done with NT-MDT, Ntegra Prima AFM instrument. The silicon cantilever with radii of 10 nm was used in semi-contact mode. The scan image was 256x256 points.

X-ray photoelectron spectra (XPS) of the brush structure were measured with using a SPECS-XPS (Germany) equipped with a hemispherical 9 channels analyzer and an Al K$\alpha$ monochromatic X-ray source. Survey spectra of the samples were scanned in the energy range of 0 eV − 1200 eV using an energy step of 1 eV and a pass energy of 30 eV. High resolution scans for $C_{1s}$, $N_{1s}$ and $O_{1s}$ were recorded with an energy step of 0.03 eV and a pass energy of 23 eV. Peak analysis was performed using CasaXPS software. The atomic fraction in the coatings was determined by calculating the integrated areas of the $C_{1s}$, $N_{1s}$, and $O_{1s}$ peaks with a Shirley background. Charge compensation was applied in all spectra by assigning the C−C/C−H component in the $C_{1s}$ peak at a binding energy of 285 eV.

## 3. Results

*3.1. Characterisation of the carbonised substrates*

The PS spun layer was converted to carbonized substrate with PIII treatment. The hydrophilicity of the surface was significantly increased after PIII. The water wetting angle decreased from 90º for the untreated PS to 60º for the carbonized PS.

The surface topography of the spincoated polystyrene film is similar to the topography of the silicon substrate itself. The polystyrene surface topography was not changed after PIII as observed in AFM image (Fig.2a). The Root Mean Square (RMS) roughness for the carbonized layer (0.257 nm) is comparable to the RMS roughness of the silicon wafer. The height histogram is fitted by the single Gauss function with the width of 0.57 nm.

The Raman spectrum of the carbonized substrate (Fig.2c) shows intense, wide peaks at 1543 and 1384 cm$^{-1}$. The lines of the polystyrene vibrational modes are not observed. The 1543 cm$^{-1}$ peak corresponds to the $E_{2g}$ vibration mode and 1384 cm$^{-1}$ peak corresponds to the $A_g$ vibrational mode of a graphitic ring. Following the model of Raman spectra for carbon structures, the observed spectrum corresponds to glassy carbon. According to the position of the G-peak at 1546 cm$^{-1}$ and the $I_D/I_G$ ratio of 0.65, the carbonized structure contains nanocrystalline graphitic clusters of about 1 nm in size with sp$^3$ hybridized of carbon atoms on the edges of the clusters.

The refractive index at 500 nm wavelength of light measured by ellipsometry increases from 1.6 for untreated PS to 2.4 for the carbonized layer. Such a high refractive index corresponds to a dense carbon structure [17].

ESR spectra of the carbonised PS show a peak with g-factor at 2.0025 that corresponds to the unpaired electrons at the edges of the graphitic clusters (Fig.2d). A high intensity of the peak remains after long storing time (month).

The FTIR transmission spectrum obtained from the untreated spincoated polystyrene films corresponds to that obtained from bulk polystyrene but with lower intensity peaks due to the thickness being only 100 nm. As shown in Fig.2e, the spectrum shows lines due to aromatic ring vibrations at 3081, 3060, 3026, 1602 and 1493 cm$^{-1}$ and due to the aliphatic backbone of the polystyrene macromolecule at 2922, 2851 and 1452 cm$^{-1}$. The intensities of the lines decrease after PIII treatment. In addition to the decrease in intensity observed in the characteristic polystyrene peaks, a new broad band is observed in the 1750-1600 cm$^{-1}$ region. This band has a complex shape arising from overlapping lines attributable to ν(C=O) stretch vibrations in carbonyl, carboxyl, aldehyde and ester groups as well as ν(C=C) stretch vibrations in new different unsaturated carbon-carbon groups. The appearance of these new lines is due to the dehydration of the PS macromolecules and restructuring to the unsaturated crosslinked fragments, and oxidation of the carbon containing free radicals in atmosphere. Atmospheric oxidation after treatment also creates a weaker broad band in of the 3600-3400 cm$^{-1}$ spectral region attributed to ν(O-H) stretch vibrations in hydroxyl, carboxyl and hydroperoxide groups.

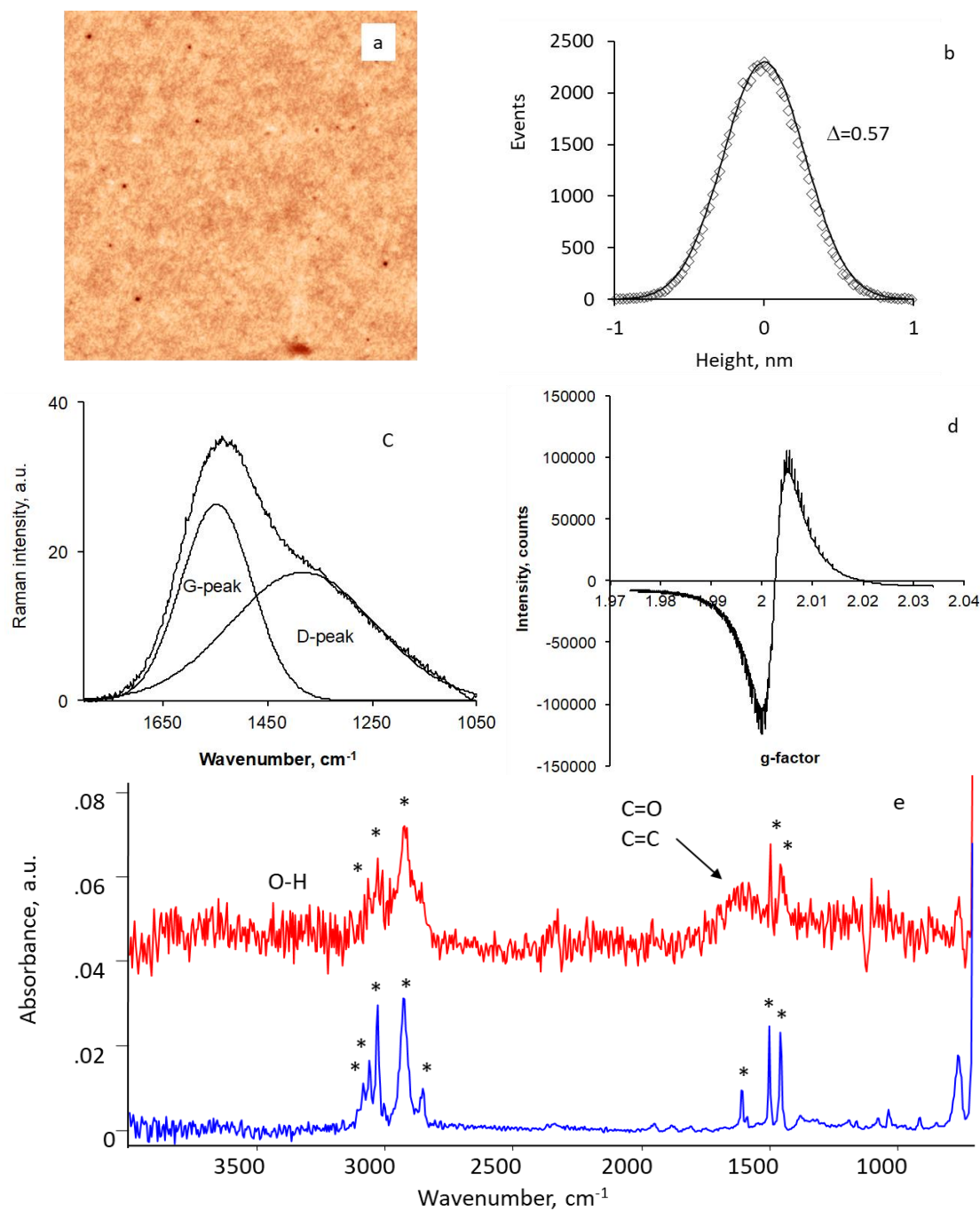

Fig.2. Carbonized substrate. (a) AFM topography image of 5x5 μm size and (b) its statistical analysis of carbonized PS on silicon wafer. (c) Raman spectrum of carbonized PS shows D and G peaks. (d) Electron spin resonance spectrum of the carbonized PS shows a signal of unpaired electrons. (e) FTIR transmission spectra of untreated PS (bottom, blue) and carbonized PS (top, red). The red spectrum is zoomed in 4 times. The spectra of silicon wafer and baseline are subtracted.

3.2. Characterisation of the brush structures

The brush attachment without linker on the carbon coating after PIII was not done before. We did not find any references on the attachment process. Because of this, we have used the attachment process developed for the linker chemistry. In order to optimise the attachment process without linker, the attachment conditions have been investigated on an example of the brush polymer based on PS chain with one thiol group.

*3.2.1. The polymer with one thiol group (P4434, PS-SH, $M_n$=50000) was spun and annealed for 17 hours at 120°C.*

The surface of brush structure is smooth as a surface of the carbonized layer (Fig.3). The phase image is smooth that shows uniformly distributed brush polymer. The topography histogram is fitted by the single Gauss function peak, which has a width of 0.57 nm. This histogram width equals to the carbon coating histogram width of 0.57 nm. The fitting of the ellipsometry data shows 6.4±1.2 nm of average thickness of the dry brush layer on the top of the carbonized coating. This thickness gives 0.08±0.01 nm$^{-2}$ value for the grafting density of the polymer. The reduced tethered density calculated by (2) from the grafting density is 3.4±0.7 that corresponds to a transition regime of the grafted polymer.

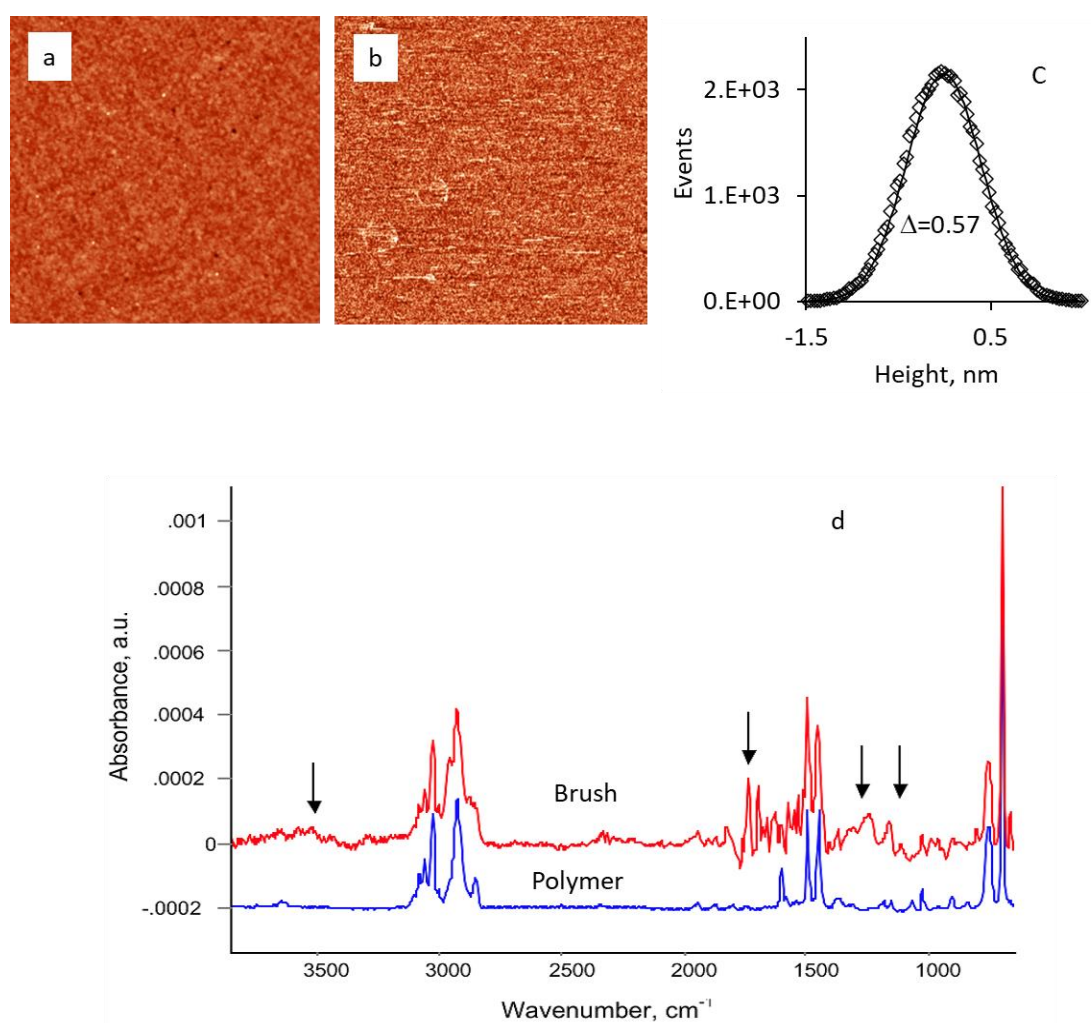

Fig.3. (a) AFM topography of 5x5 μm area, (b) phase image and (c) statistical analysis of P4434 (PS-SH, $M_n$=50000) brush polymer spun and annealed, (d) FTIR transmission spectrum of the attached brush structure (red, top) and the bulk brush polymer (blue, bottom) The spectra of silicon wafer and the carbonised layer are subtracted. Arrays show lines discussed in text.

FTIR transmission spectra show a presence of the brush polymer (Fig.3). The spectrum of thick polymer layer contains the characteristic lines of the polystyrene backbone of the polymer brush molecule. The same lines of lower intensity are observed in the spectra of attached polymer on the carbonised substrate. The clear visible peaks of PS macromolecule are observed in high frequency region at 3081, 3060, 3026, 2922 and 2851 cm$^{-1}$. The peaks at 1602, 1493 and 1452 cm$^{-1}$ are observed, but disturbed with $\nu$(C=O) peaks at 1700-1750 cm$^{-1}$ region. The narrow spikes attributed to water vapour in the spectrometer cannot be fully subtracted in this range of absorbance. The absorbance of the PS peaks corresponds to the thickness of the brush layer about 6.4 nm.

The attachment of the same brush polymer at room temperature without annealing did not give any grafted polymer layer. It is observed by ellipsometry, AFM and FTIR spectra, which show bare carbon substrate.

The attachment on the fresher carbon coating in 3 weeks after PIII with annealing at 120ºC gave the thicker layer of the brush (9.47 nm). This thickness gives 0.12±0.01 nm grafting density of the brush polymer. The reduced tethered density calculated by (2) from the grafting density is 5.1±0.7 that corresponds to a brash regime of the grafted polymer. This brush was swollen in toluene and measured with ellipsometry. The thickness of the swollen brush is 260 nm. This thickness is in 1.8 times higher than the length of the grafted molecule.

*3.2.2. The polymer with one thiol group (P4434, PS-SH, $M_n$=50000) is attached from a wet drop of THF solutions during overnight at room temperature.*

One of the methods to attach brush polymer is an attachment from solution. THF solutions of 1 and 10% concentrations were used. In this method the attachment was done at room temperature without annealing. The surface with the grafted polymer is rougher than the carbonized layer (Fig.4). The AFM phase image shows a presence of two materials: first material is the hard carbonized bottom layer and second layer is the soft brush polymer. The soft material forms the features higher than the hard material layer. The topography histogram is fitted with two Gauss functions. The first peak has a width (0.52 nm), which is similar to the carbonized surface (0.57 nm). The second peak has higher width (1.3 nm) related to the grafted polymer. Power Spectral Density analysis (PSD) shows 28, 113 and 2500 nm characteristic dimension of the polymer features. Ellipsometry data of the dry grafted polymer were fitted with 1.86 nm of the averaged thickness. This averaged thickness of the polymer gives 0.023 nm$^{-2}$ of grafting density calculated by formula (1). The reduced tethered density calculated by formula (2) is 1.0 that corresponds to a mushroom regime of the grafted polymer. Therefore, the attachment from 1% solution at room temperature is not enough to form a brush structure.

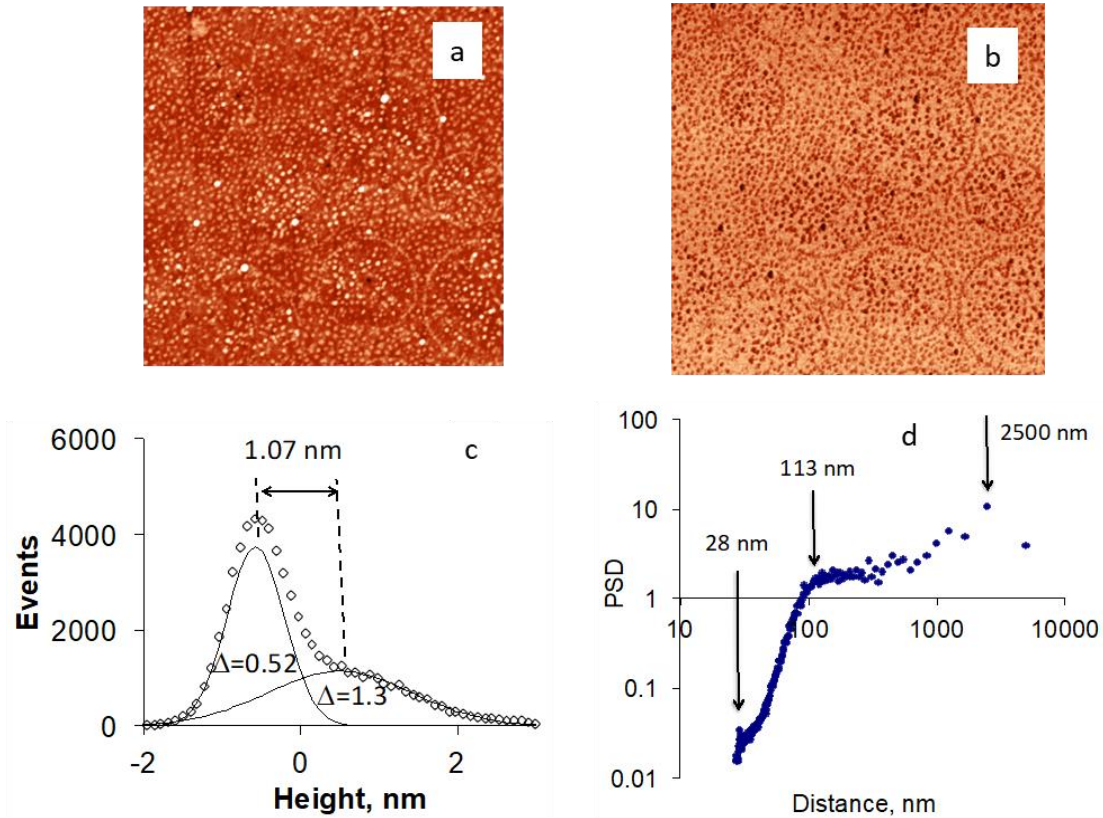

Fig.4. (a) AFM topography of 5x5 μm area, (b) phase image and (c) statistical analysis of P4434 (PS-SH, $M_n$=50000) brush polymer attached from THF 1% solution.

The surface with the grafted polymer is much rougher than the carbonized layer (Fig.5). The AFM phase image shows a presence of two materials: first material is the hard carbonized bottom layer and second layer is the soft brush polymer features higher than the bottom layer. The topography histogram is fitted with four Gauss functions: bottom surface of 0.76 nm width corresponded to the width of carbonised substrate, the 2 nm height structures of 1.6 nm width, 4 nm height structures of 0.74 nm width, and 5 nm height structures of 1 nm width. The phase image shows two kinds of materials. The FTIR transmission spectrum shows the lines attributed to PS macromolecule as well as C=O lines. Additionally, the $CH_2$ group vibrational lines and C-O vibrational lines related to THF are observed. The ellipsometry data were fitted with 2.15±0.26 nm of the average thickness of the grafted layer. The averaged thickness gives 0.027±0.003 $nm^{-2}$ grafting density of the polymer. The reduced tethered density calculated from the grafting density is 1.1±0.1 that corresponds to a mushroom regime of the grafted polymer. Therefore, the attachment from 10% solution at room temperature is also not enough to form a brush structure.

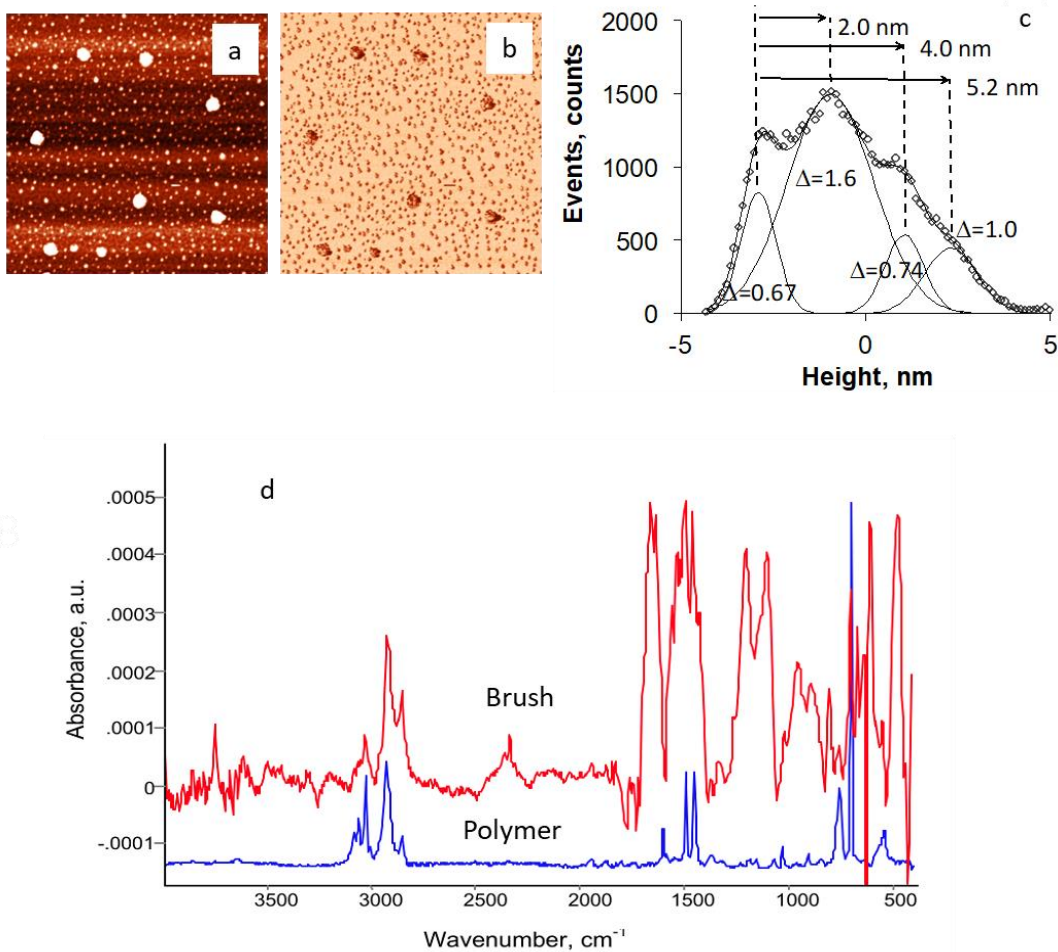

Fig.5. (a) AFM topography of 5x5 μm area, (b) phase image and (c) statistical analysis of P4434 (PS-SH, $M_n$=50000) brush polymer attached from THF 10% solution, (d) FTIR transmission spectrum of the attached brush structure (red, top) and the bulk brush polymer (blue, bottom) The spectra of silicon wafer and the carbonised layer are subtracted.

*3.2.3. The polymer with one amine group (P3702, PS-NH$_2$, $M_n$=32000) is attached from spun layer with annealing overnight at 120°C.*

The surface of the grafted polymer is viewed smooth as the carbonized coating (Fig.6). The phase image is smooth that shows uniformly distributed grafted polymer on the surface. The topography histogram is fitted with the single Gauss function of 0.47 nm width and lower than 0.57 nm width of the carbonised substrate. The ellipsometry measurements are fitted with 6.2±1.1 nm average thickness of the dry polymer layer. The grafted density calculated from the ellipsometry results is 0.12±0.02 nm$^{-2}$. The reduced tethered density is 3.3±0.6 that corresponds to the transition regime of the grafted polymer layer. The FTIR transmission spectra show a presence of characteristic lines for PS macromolecule.

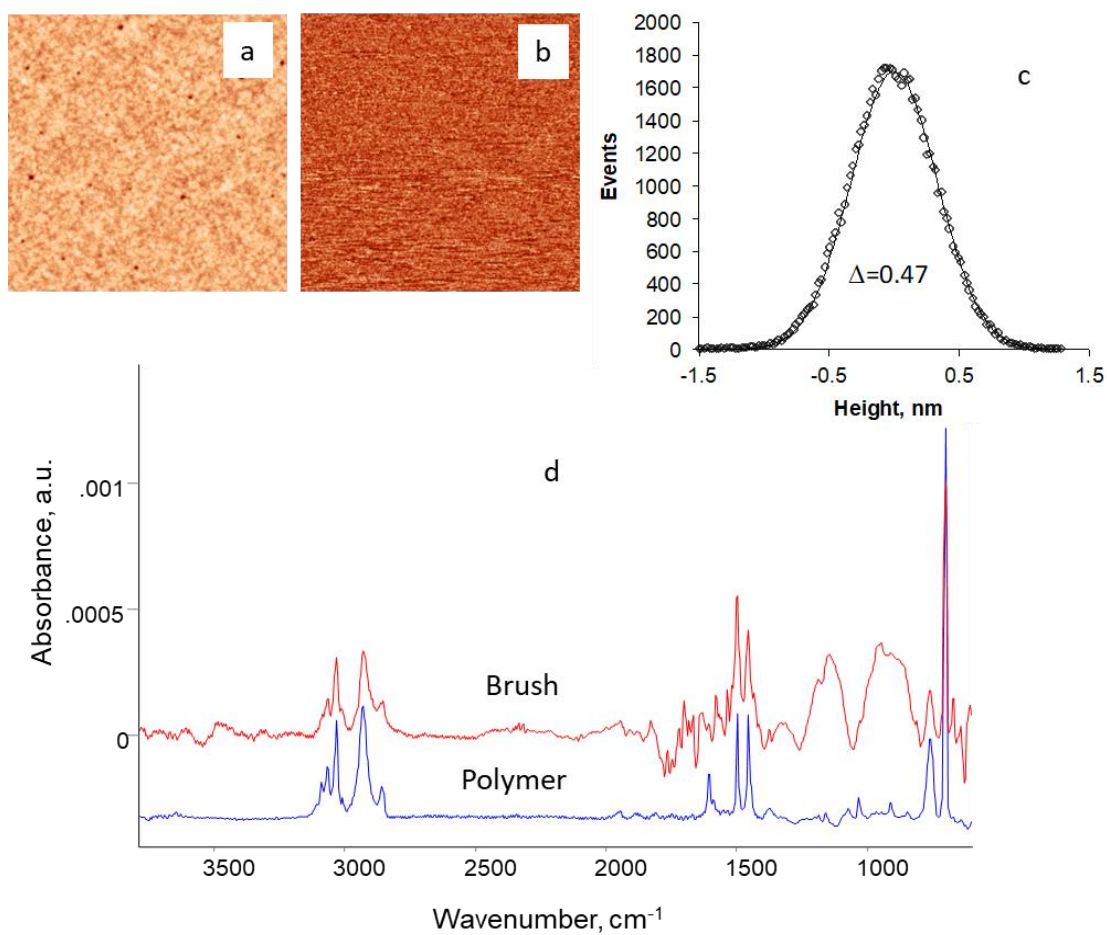

Fig.6. (a) AFM topography image and (b) phase image of 5x5 μm size and (c) its statistical analysis of P3702 (PS-NH$_2$, M$_n$=32000) brush polymer, and (d) FTIR transmission spectra of the brush and thick polymer layer.

The brush was swollen in toluene. The thickness of the swollen brush measured by ellipsometry is about 500 nm. However, the intensity of the optical signal from the brush-toluene interface was week, the result of the optical modelling thickness was doubted.

*3.2.4. The polymer with thiol and amine groups (P4055, HS-PS-NH$_2$, M$_n$=39000) is attached from spun layer with annealing overnight at 120ºC.*

The surface of the grafted polymer is smooth (Fig.7). The phase AFM image shows only one material uniformly distributed on the surface. The topography histogram is fitted with single Gauss function of 0.57 nm width, that equals to the carbonised substrate. The ellipsometry data shows 5.0±0.1 nm thickness of the dry grafted polymer layer. The grafting density calculated from this thickness is 0.08 nm$^{-2}$. This grafting density corresponds to 2.7 reduced tethered density that means the transition regime of the grafted polymer layer.
FTIR spectrum shows the peaks associated with the vibrations in PS macromolecule (Fig.7). Minor oxidation peaks are observed in 1700-1750 cm$^{-1}$ region.

The ellipsometry measurements of the grafted polymer in toluene are fitted with the swollen brush of 106 nm thickness. This value is close to the 112 nm length of the brush polymer molecule calculated from the molecular weight and geometry of the molecule.

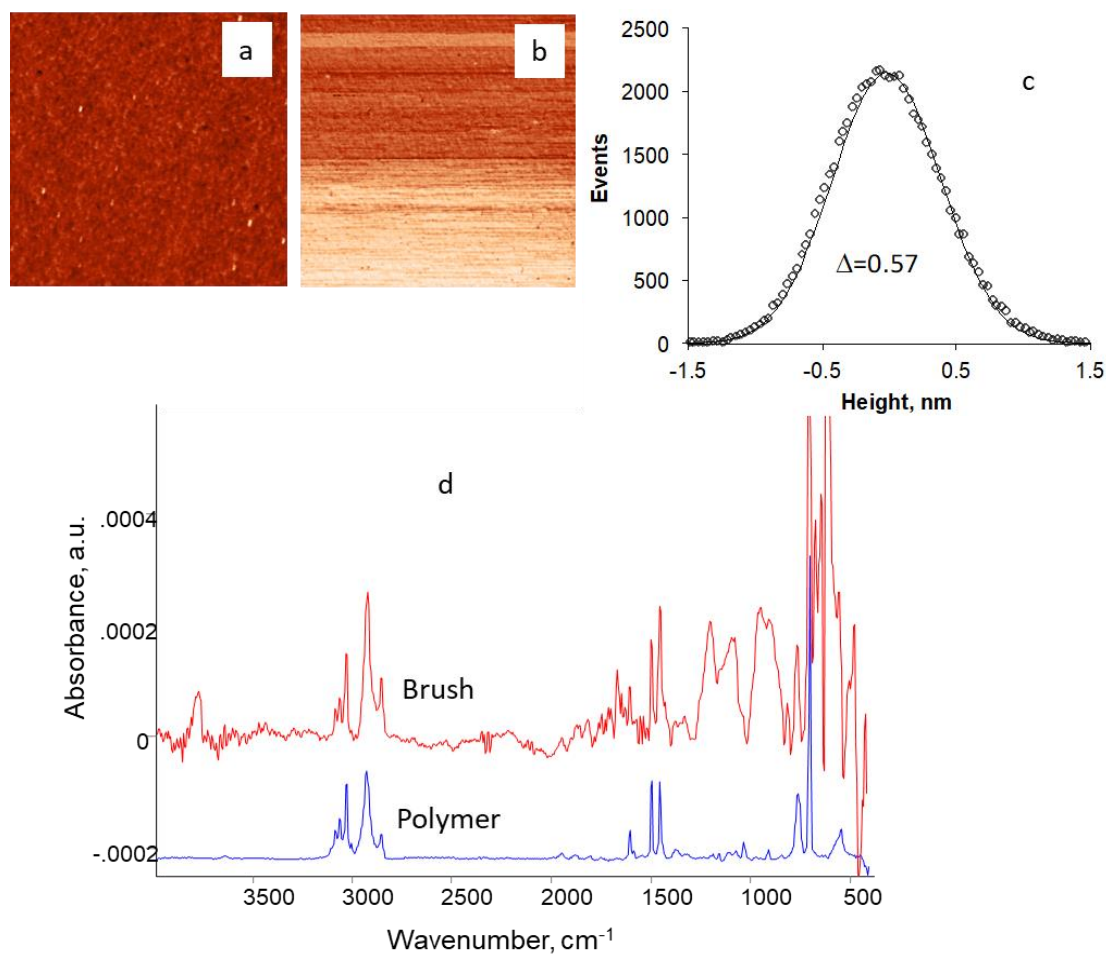

Fig.7. (a) AFM topography image and (b) phase image of 5x5 μm size and (c) its statistical analysis of P4055 (HS-PS-NH$_2$, M$_n$=39000) brush polymer, and (d) FTIR transmission spectra of the brush and thick polymer layer.

*3.2.5. The polymer with thiol and amine groups (P4041C, HS-PS-NH$_2$, M$_n$=553000) attached from spun layer with annealing overnight at 120ºC.*

The surface of the dry brush is smooth (Fig.8). The topography histogram is fitted with single Gauss function of 0.45 nm width. The phase AFM image shows a uniform distribution of a single kind of material over the whole surface. The ellipsometry data are fitted with 12.6 nm thickness of the dry grafted polymer layer. The grafting density calculated from the ellipsometry data is 0.015 nm$^{-2}$ that is relatively low value. However, the corresponding reduced tethered grafting density is 6.9, that means the grafted polymer is in a brush regime. FTIR spectrum shows the peaks associated with the vibrations in PS macromolecule. Some broad peak in 1650-1750 cm$^{-1}$ region interpreted as the oxidation is observed.

The thickness of the swollen brush in toluene is 240 nm, while the length of the molecule calculated from the molecular weight and geometry of the molecule is 1586 nm.

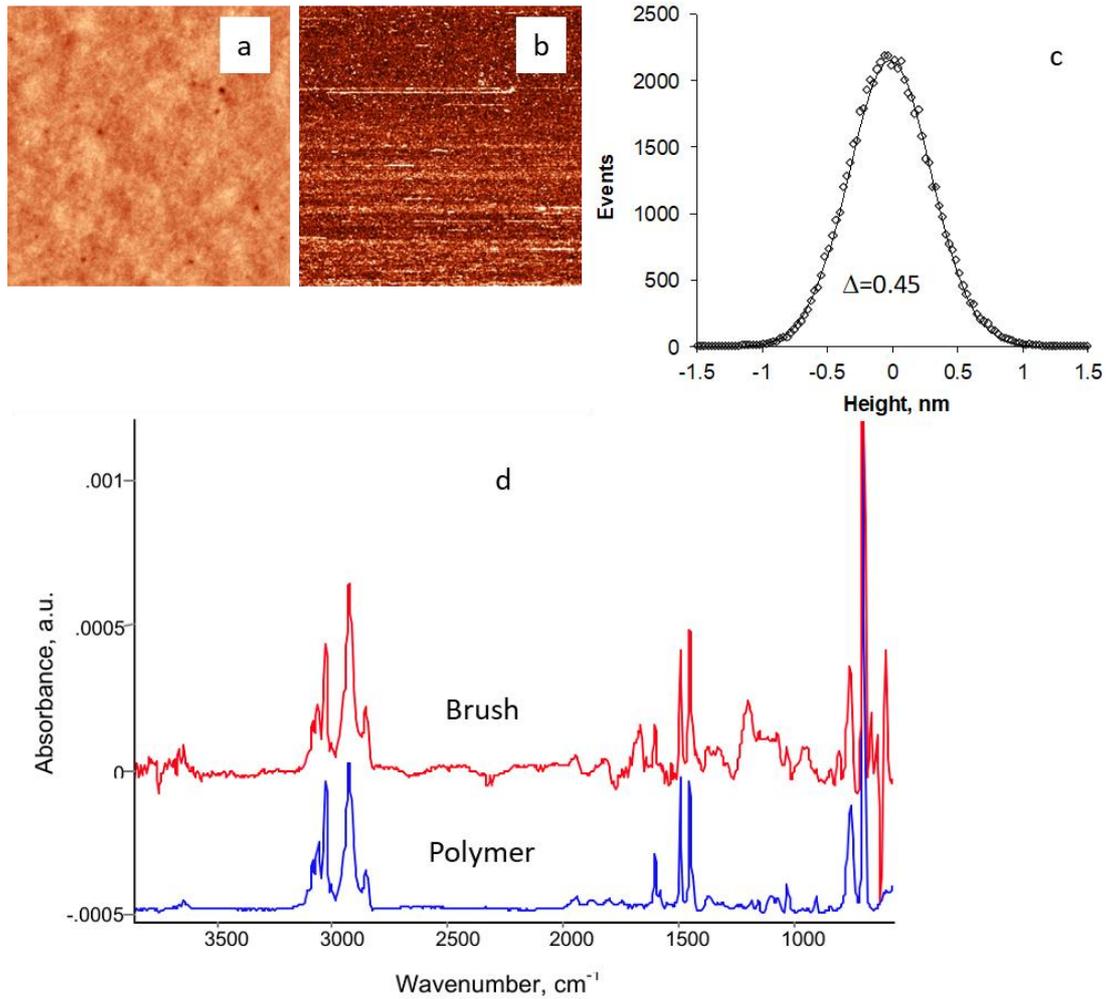

Fig.8. (a) AFM topography image and (b) phase image of 5x5 µm size and (c) its statistical analysis of P4041C (HS-PS-NH$_2$, M$_n$=553000) brush polymer, and (d) FTIR transmission spectra of the brush and thick polymer layer.

*3.2.6. The polymer with one carboxyl group (P2824, PS-COOH, M$_n$=48000) attached from spun layer with annealing overnight at 120°C.*

The surface of the dry brush is smooth (Fig.9). The AFM phase image shows the grafted polymer covers uniformly the surface. The ellipsometry measurements show 3.6±0.3 nm of the average thickness of the dry grafted layer. FTIR transmission spectra show a presence of PS backbone of the grafted polymer. The grafting density calculated from the ellipsometrical data is 0.05±0.04 nm$^{-2}$. The corresponding reduced tethered grafting density is 1.9±0.2, that means the grafted polymer is in a transitional regime.

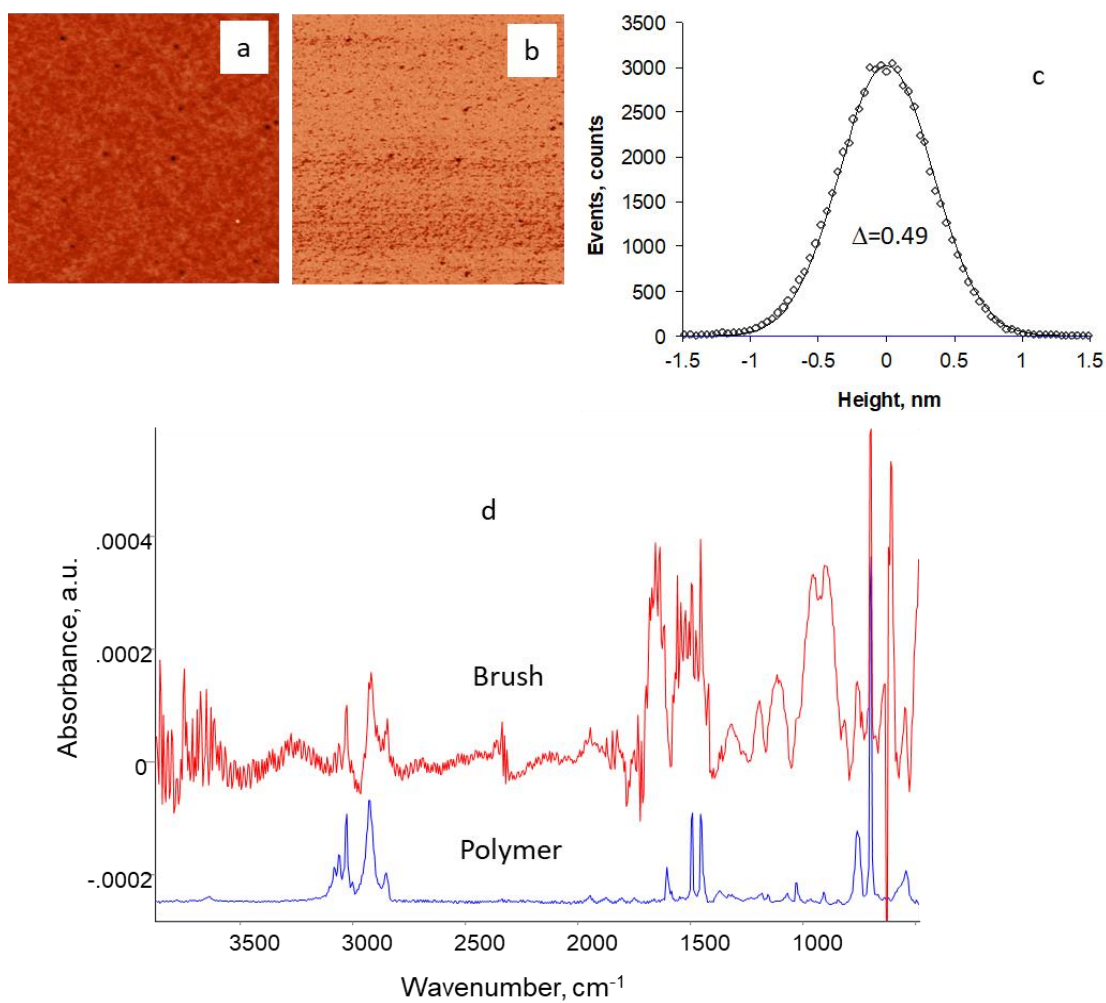

Fig.9. (a) AFM topography image and (b) phase image of 5x5 μm size and (c) its statistical analysis of P2824 (PS-COOH, $M_n$=48000) brush polymer, and (d) FTIR transmission spectra of the brush and thick polymer layer.

*3.2.7. The polymer with thiol and amine groups (P4030, HS-PS-NH$_2$, $M_n$=21500) attached from spun layer with annealing overnight at 120ºC.*

The topography of the brush is complicate (Fig.10). The topography histogram is fitted with minimum 3 Gauss functions. The bottom function has a width of 0.71 nm, that corresponds to the carbon layer surface. This carbon layer covers about 1.2 % of the total area. The middle Gauss function corresponds to the grafted polymer surface with a width of 1.3 nm. This layer is a surface of the grafted polymer and covers 90.1 % of the total area. The top Gauss function has similar width of 1.2 nm and covers 8.7 % of the total area. Power Spectral Density analysis of the AFM topography gives the features of 10, 40, 90 and 330 nm sizes. The phase image shows minimum two different materials: one material is on the tops of the hills and the second material in the valleys. The thickness of dry brush is 15.2 nm by ellipsometry data.

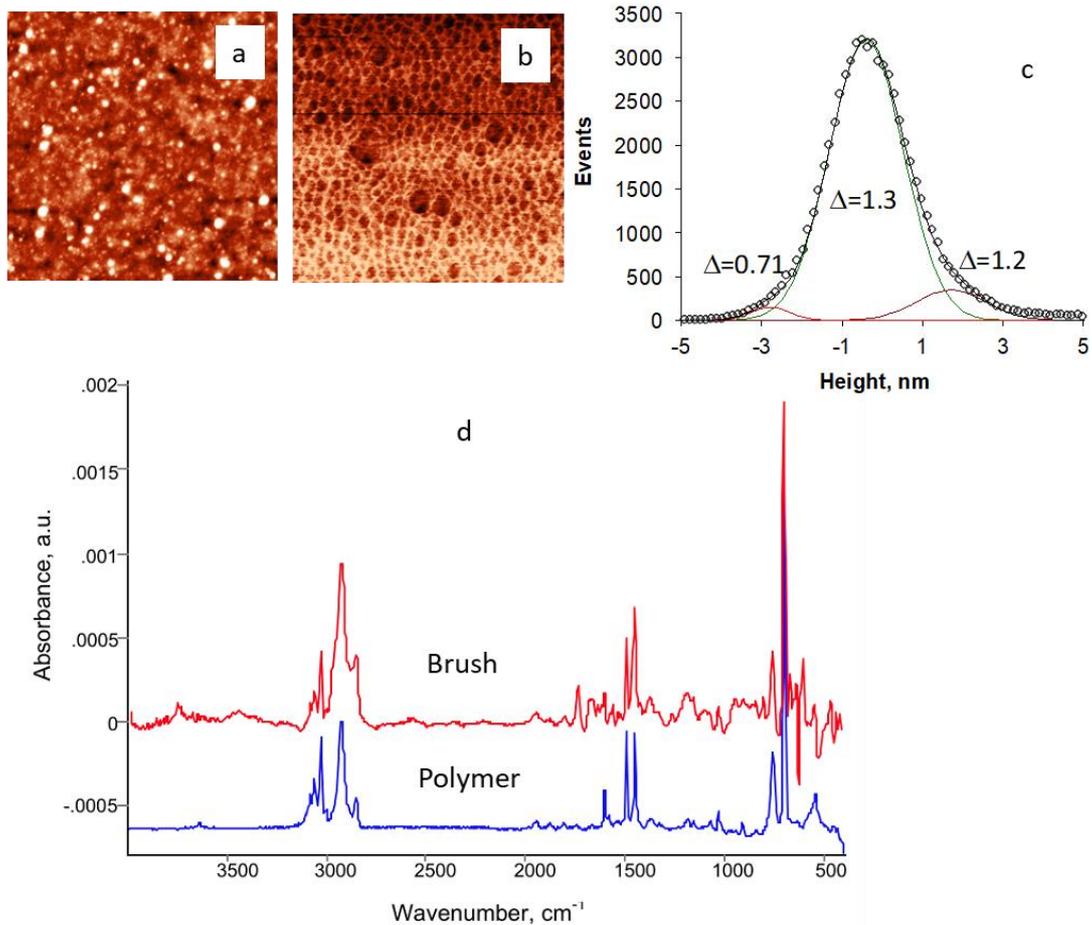

Fig.10. (a) AFM topography image and (b) phase image of 5x5 μm size and (c) its statistical analysis of P4030 (HS-PS-NH$_2$, M$_n$=21500) brush polymer, and (d) FTIR transmission spectra of the brush and thick polymer layer.

XPS shows (Fig.11) higher concentration of the sulphur peak in the initial bulk polymer (2.3%), than the amount calculated by formula (0.017%). The brush structure has 3.1% of sulphur, that is higher, than in the initial polymer and calculated by formula. The amount of oxygen (3.6%) and nitrogen (0.033%) in the bulk polymer is higher than the amount of oxygen (0%) and nitrogen (0.017%) calculated by chemical formula. The amount of oxygen (2.3%) and nitrogen (0.069%) in the spectra of the brush structure does not correspond to the theoretical values also. FTIR spectrum shows the vibration peaks of the PS macromolecules. The additional peaks of ν(OH) and ν(C=O) vibrations indicate about minor oxidation of the brush structure. When the brush is swollen in the toluene, the thickness is 101 nm. The length of the molecule calculated from the molecular weight and geometry of the molecule is 62 nm.

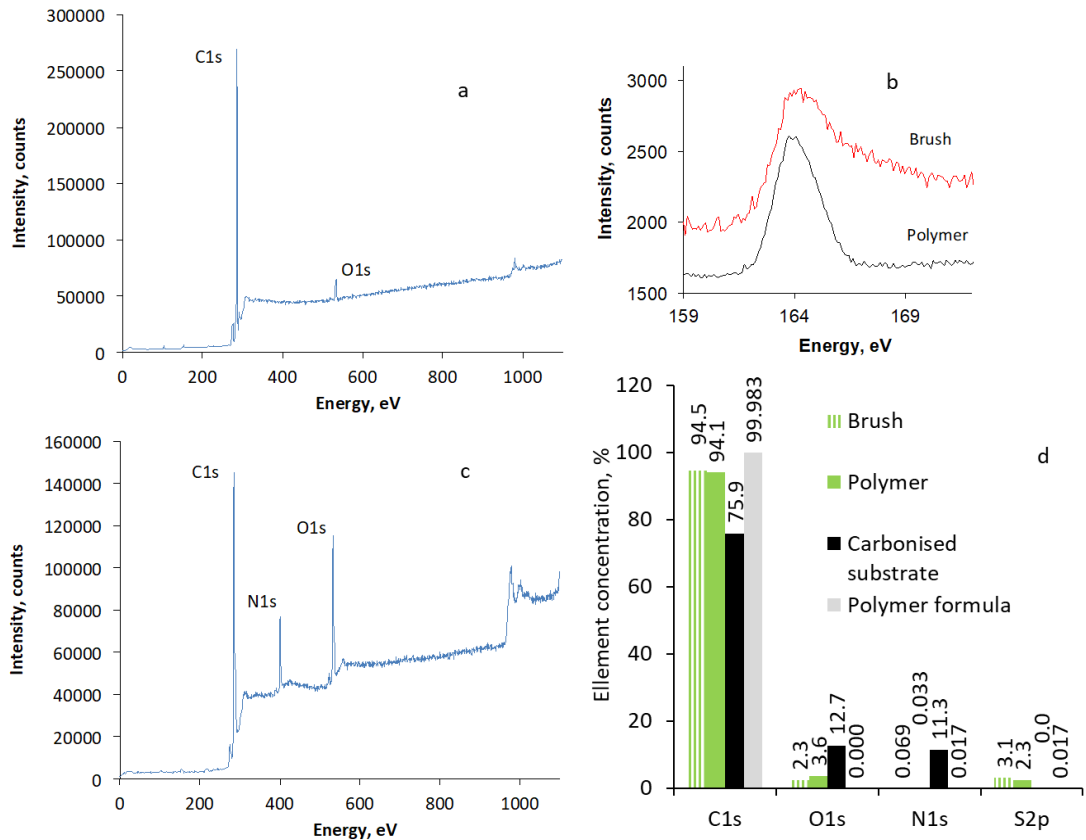

Fig.11. (a) XPS spectra of P4030 brush and (c) carbonized PS on silicon wafer before a brush attachment. (b) XPS $S_{2p}$ line of P4030 polymer as a powder and as a brush. (d) Element content of P4030 polymer in brush and bulk powder, as well as theoretical values following chemical formula, and carbonized substrate for comparison.

*3.2.8. The brush with one amine groups (P6145A, NIPAM-NH$_2$, $M_n$=50500) attached from spun layer with annealing overnight at 150°C.*

The brush surface is smooth (Fig.12). The phase image shows uniformity over the surface. The height histogram is fitted by single Gauss function with 0.52 nm width corresponding to the 0.57 nm width of the carbonised substrate. The thickness of the dry brush is 5.7 nm. The thickness of the swollen brush in water is 106 nm, while the theoretical length of the molecule is 133 nm. The FTIR spectrum shows the peaks of acrylamide backbone such as -NH, -CH$_2$-, C=O, C-O groups.

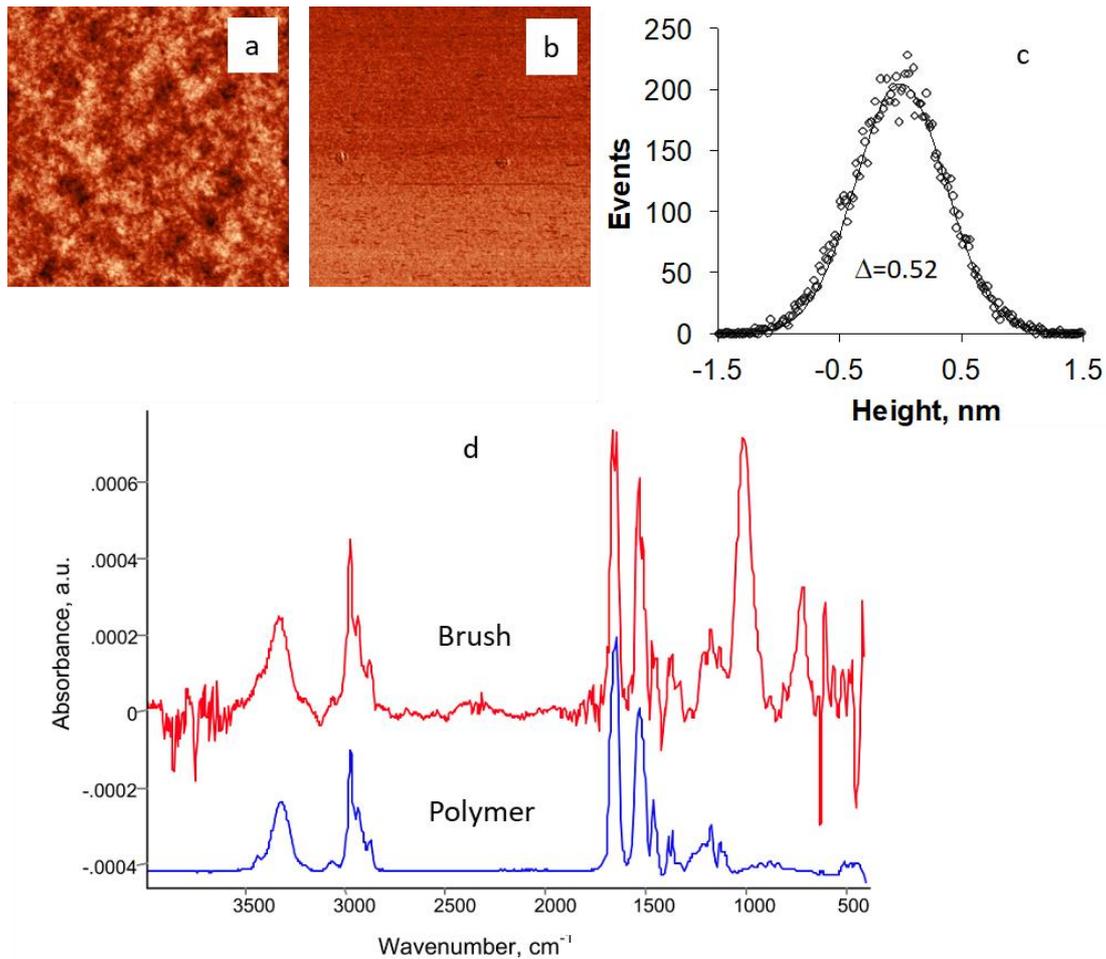

Fig.12. (a) AFM topography image and (b) phase image of 1x1 μm size and (c) its statistical analysis of P6145A (NIPAM-NH$_2$, M$_n$=50500) brush polymer, and (d) FTIR transmission spectra of the brush and thick polymer layer.

*3.2.9. The brush with one carboxyl group (P5590, NIPAM-COOH, M$_n$=45000) attached from spun layer with annealing overnight at 150ºC.*

The brush surface is smooth (Fig.13). The phase image shows uniform top layer. The height histogram is fitted by single Gauss function with width of 0.74 nm. This is slightly higher than the 0.57 nm width for carbonised substrate surface. The thickness of the dry brush is 3.4 nm. The thickness of the swollen brush in water is 48 nm, when the theoretical length of the molecule is 119 nm. The FTIR spectrum shows some peaks attributed to -NH, -CH$_2$-, C=O, C-O group vibrations in acrylamide backbone.

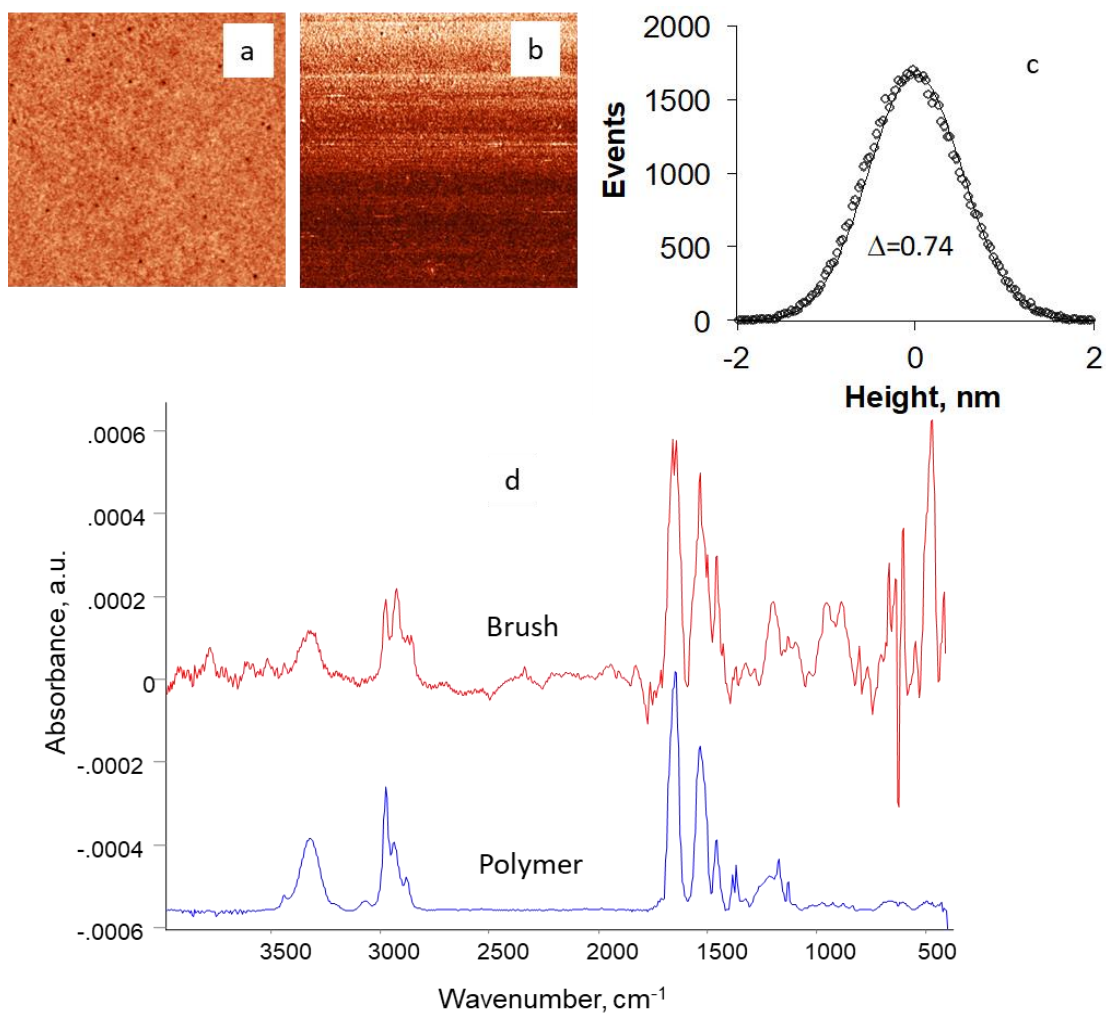

Fig.13. (a) AFM topography image and (b) phase image of 5x5 μm size and (c) its statistical analysis of P5590 (NIPAM-COOH, $M_n$=45000) brush polymer, and (d) FTIR transmission spectra of the brush and thick polymer layer.

*3.2.10. The brush with thiol and carboxyl groups (P6698, HS-NIPAM-COOH, $M_n$=30000) attached from spun layer with annealing overnight at 150ºC.*

The brush surface is rough (Fig.14). The features of 8, 100 and 300 nm sizes are observed. However, the phase image shows sufficient uniformity of the attached layer. That means, the brush polymer coats the surface completely, but the thickness of the dry layer is not uniform. The height histogram is fitted by minimum two Gauss functions. The average thickness of the dry brush is 3.8 nm. The thickness of the swollen brush in water is 154 nm, while the length of the molecule is 79 nm. The FTIR spectrum shows the peaks of acrylamide backbone: -NH, -$CH_2$-, C=O, C-O peaks, however, the intensity of the spectral peaks is different, than the peaks in the spectra of the initial brush polymer. Additional strong line at 1000 $cm^{-1}$ is observed.

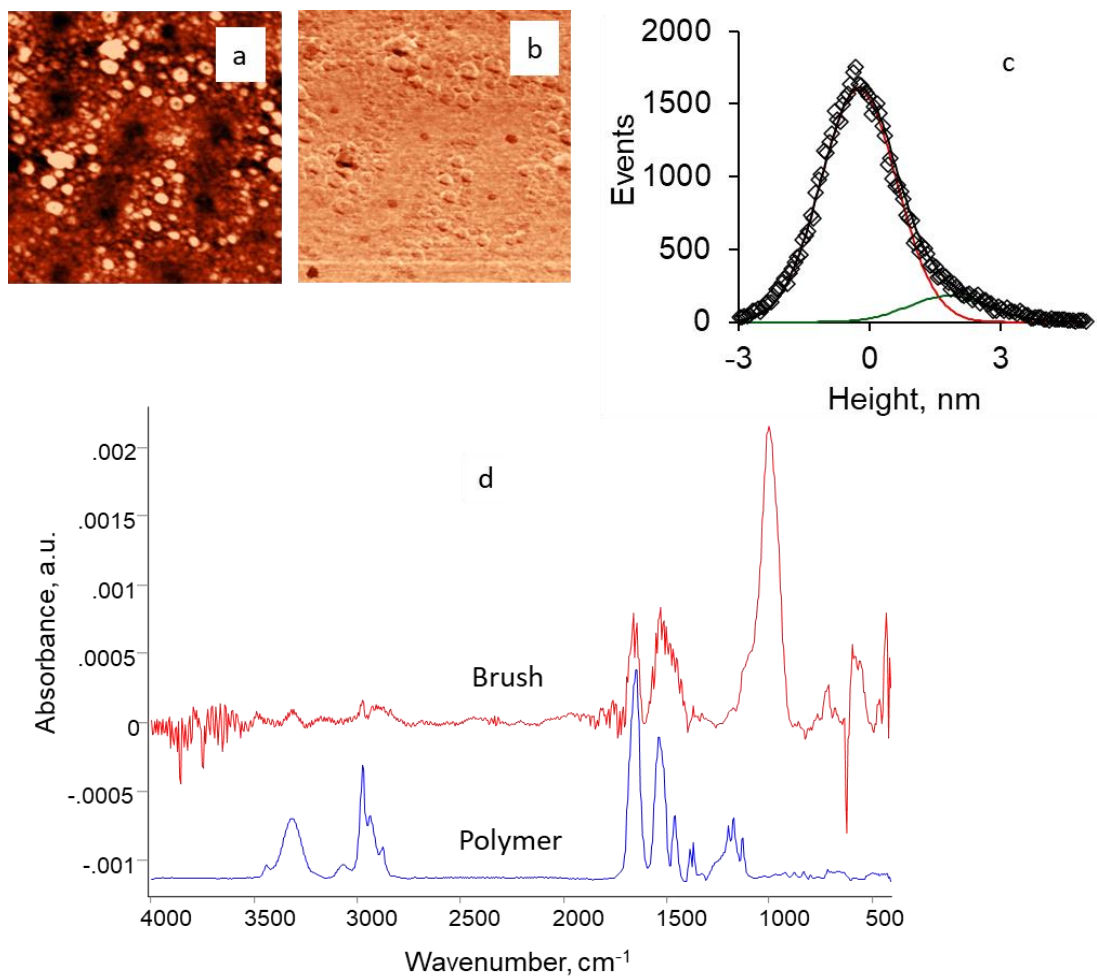

Fig.14. (a) AFM topography image and (b) phase image of 1x1 μm size and (c) its statistical analysis of P6698 (HS-NIPAM-COOH, $M_n$=30000) brush polymer, and (d) FTIR transmission spectra of the brush and thick polymer layer.

## 3.3. Swelling of the brush structures

The brush structures based on acrylamide backbone have been tested by ellipsometry in the liquid cell with water under temperature range from 20 to 50ºC. The cell was heated and cooled in reverse regime.

The thickness and the refractive index at 632.8 nm wavelength of the brush polymer P6145A (NIPAM-NH$_2$, $M_n$=50500) at cycles of the heating and cooling is shown in Fig.15. The initial brush thickness in dry state is 4.48 nm. Phase transition in the bulk polymer is observed at 30ºC. The phase transition in the brush structure is observed between 30 and 32$^0$C. The thickness at the phase transition is collapsed from 120 nm to 40 nm. When temperature is achieved 50ºC and higher, the thickness goes higher again. The refractive index drops from 1.3320 to 1.3295 in the same temperature range and recovers with following temperature rise. The changes repeat at the second cycle of the heating and cooling.

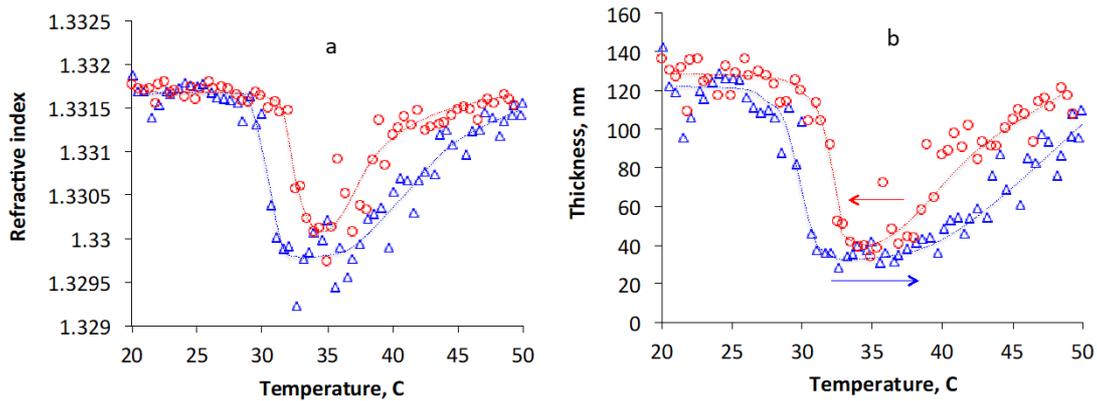

Fig.15. Refractive index (a) and thickness (b) of P6145A brush under water in dependence on temperature by ellipsometry data. Heating (triangle, blue) and cooling (circle, red) stages.

The thickness and the refractive index at 632.8 nm wavelength of the brush polymer P5590 (NIPAM-COOH, $M_n$=45000) at cycling heating and cooling is shown in Fig.16.

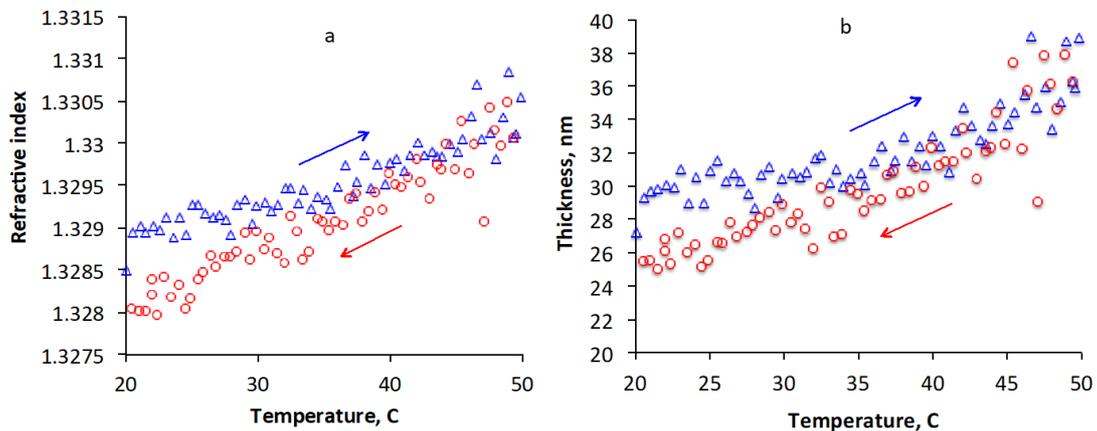

Fig.16. Refractive index (a) and thickness (b) of P5590 brush under water in dependence on temperature by ellipsometry data. Heating (triangle, blue) and cooling (circle, red) stages.

The thickness of the brush increase from 28 nm to 37 nm and the refractive index increases from 1.329 to 1.3305 with temperature increase from 20 to 50ºC. When temperature decrease, the thickness decrease to 26 nm and refractive index decreases to 1.328. No phase transition in the brush structure is observed.

The thickness and the refractive index at 632.8 nm wavelength of the brush polymer P6698 (HS-NIPAM-COOH, $M_n$=30000) at cycling heating and cooling is shown in Fig.17. The thickness of the brush increase from 54 nm to 64 nm and the refractive index increases from 1.3315 to 1.3319 with temperature increase from 20 to 50ºC. When temperature decrease, the thickness decrease to 54 nm and refractive index decreases to 1.3313. No phase transition in the brush structure is observed.

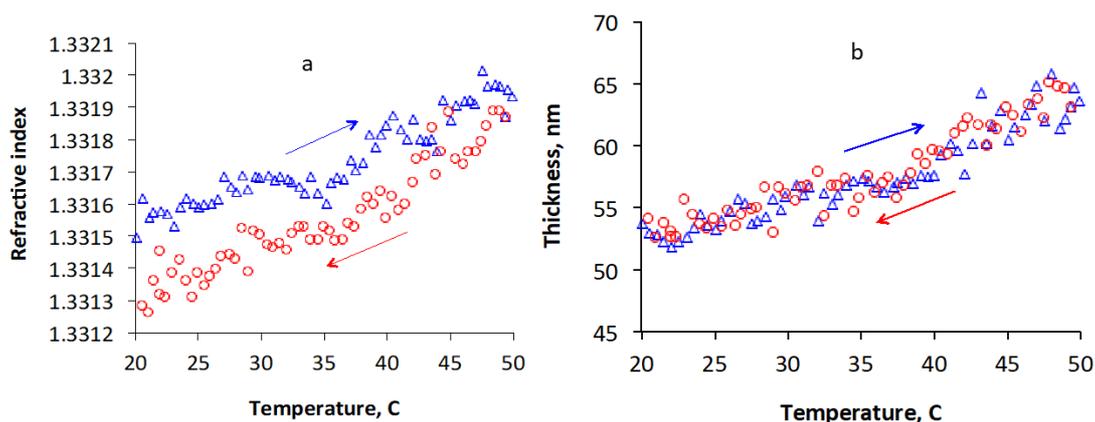
Fig.17. Refractive index (a) and thickness (b) of P6698 brush under water in dependence on temperature by ellipsometry data. Heating (triangle, blue) and cooling (circle, red) stages.

## 4. Discussion

Brush structures have the potential to control the properties of the surface to which they are grafted. The linker chemistry in some cases is impossible or difficult the use of brush structures due to the complexity of the technological process, as well as the toxicity of the components, if this is further used in biomedical devices. Therefore, obtaining surfaces with brush structures without linkers is of interest.

In this work, the polymer surface activity obtained by ion-beam treatment was used to attach brush structures without linkers. As a result of the treatment, the polymer surface undergoes deep structure changes and condensed aromatic carbon clusters with unsaturated valences at the edges of the clusters at the carbon atom, or, in other words, with free radicals, are formed. Such free radicals retain activity for a long time after treatment due to the stabilization of the unpaired electron on the π-electron cloud of the aromatic cluster. On the other hand, if other substances are now applied to the treated surface, then, when interacting with the molecules adsorbed on the surface, free radicals can react to form a covalent bond. As a result of the reaction, a strong crosslink is formed between the polymer surface and the adsorbed molecules. This property of the treated surface to link the molecules was used to obtain a brush structure. However, the universality of the free radical activity to almost any chemical group of adsorbed molecules did not allow predicting the possibility of successfully obtaining a brush structure. A molecule in a brush structure must be attached to the surface with only one active end. Attaching a molecule with two ends or via a random group in the middle of the molecule will disrupt the brush structure.

The results of this work show that the application of brush molecules from a solution, as is usually done in the linker technology, does not lead to a continuous coating of the surface with molecules. The coating is patching and the brush structure shows the mushroom mode. This corresponds to the theory of the formation of a covalent bond by reaction with stabilized free radicals. To form a covalent bond, it is necessary for the molecule to be adsorbed on the surface with some resident time.

The application of brush structures by the spin coating method and subsequent annealing gave a significantly greater effect. For a number of molecules, dense layers of brush structures with a reduced tethered density of up to 8.3 units were obtained. It should be noted that positive results were mainly observed for molecules with terminal amine and mercaptan groups. Molecules with carboxyl terminal groups formed much more rare coatings or did not form any at all. This also corresponds to the grafting reactions with free radicals, when the reaction of hydrogen abstraction from the amine and mercaptan group occurs.

It should also be noted that in some cases the coating chemical structure did not correspond to the formula of the brush molecules. Apparently, in such cases, the by-products of the synthesis reaction in the composition of the brush molecules played a role. Probably, these contaminants have a higher rate of attachment to the surface by the reaction of free radicals and closed the active centres on the surface so that the brush molecules no longer have the ability to adsorb and react. This was observed by FTIR and XPS spectra.

The brush coating based on acrylamide showed the effect of phase transition with a temperature, just as it was observed in the brush structure obtained by the linker chemistry [18].

## 5. Conclusion

For the first time, the polymer brushes have been synthesised on the carbonized coating without a linker. The attachment of the polymer brushes is based on the reaction of free radicals in the carbonised layer with the brush molecules. Due to high chemical activity of free radical, the end group chemistry of the polymer brush does not play crucial role in the attachment. However, due to the high chemical activity of the carbonized substrate, an impurity of the polymer brushes is critical, when the contamination can attach to the surface and prevent the brush attachment.

## Acknowledgements

Author thanks for support, assistance and advises Prof. Manfred Stamm, Dr. Petra Uhlmann, Dr. Klaus-Jochen Eichhorn, Dr. Ulla König, Dr. Eva Bittrich, Mr. Roland Schulze (Institute of Polymer Research, Dresden, Germany); Prof. Marcela Bilek (University of Sydney, Sydney, Australia); Prof. Hynek Biederman, Dr. Andrei Choukourov, Dr. Anna Artemenko, Mr. Iurii Melnichuk, Mr. Artem Shelemin (Charles University, Praha, Czechia); Dr. Petar Stojanov (SPECS, Berlin, Germany). The study was supported by Alexander von Humboldt Foundation.

Table. The brush polymers attached on the carbonized substrates, their molecular mass, method of attachment, grafting density (σ), reduced tethered density (Σ), thickness in dry and swollen states, and ratio of thickness in Θ-solvent and length of the molecule.

| Type and Formula | $M_n$ | Method of attachment | Dry brush | | | Wett brush | | Solvent |
|---|---|---|---|---|---|---|---|---|
| | | | Thickness, nm | σ, $1/nm^2$ | Σ | Thickness, nm | Length of molecule, nm | |
| *Polystyrene backbone* | | | | | | | | |
| P4434, PS-SH | 50000 | *Spun, 120ºC 17h | 9.47 | 0.12 | 5.1 | | | Toluene |
| P4434, PS-SH | 50000 | *Spun, 22ºC 17h | 0 | - | - | | | |
| P4434, PS-SH | 50000 | Spun, 120ºC 17h | 6.4±1.2 | 0.08±0.01 | 3.4±0.7 | 260 | 1.8 | |
| P4434, PS-SH | 50000 | 1% THF, 22ºC 17h | 1.86 | 0.023 | 1.0 | | | |
| P4434, PS-SH | 50000 | 10% THF, 22ºC 17h | 2.15±0.26 | 0.027±0.003 | 1.1±0.14 | | | |
| P3702, PS-NH$_2$ | 32000 | Spun, 120ºC 17h | 6.2±1.1 | 0.12±0.02 | 3.3±0.6 | 500 | 5.5 | |
| P4055, HS-PS-NH$_2$ | 39000 | Spun, 120ºC 17h | 5.0±0.1 | 0.08 | 2.7 | 106 | 0.95 | |
| P4041C, HS-PS-NH$_2$ | 553000 | Spun, 120ºC 17h | 12.6±0.4 | 0.015 | 6.9 | 240 | 0.15 | |
| P4030, HS-PS-NH$_2$ | 21500 | Spun, 120ºC 17h | 15.2±1.0 | 0.45 | 8.3 | 101 | 1.6 | |
| P2824, PS-COOH | 48000 | Spun, 120ºC 17h | 3.6±0.3 | 0.047±0.04 | 1.9±0.2 | | | |
| P6696, HS-PS-COOH | 3000 | Spun, 120ºC 17h | 0.58±0.09 | 0.12 | 0.3 | | | |
| *Acrylamide backbone* | | | | | | | | |
| P6145A, NIPAM-NH$_2$ | 50500 | Spun, 150ºC 17h | 5.4±1.0 | 0.07 | 2.8 | 123 | 0.93 | Water |
| P5590, NIPAM-COOH | 45000 | Spun, 150ºC 17h | 3.5±0.6 | 0.05 | 1.78 | 44 | 0.37 | |
| P6698, HS-NIPAM-COOH | 30000 | Spun, 150ºC 17h | 3.9±0.8 | 0.08 | 2.0 | 110 | 1.10 | |

*- the samples have been spun in 3 weeks after PIII treatment. Other samples have been spun in 5 weeks after PIII treatment.